\documentstyle[12pt,epsf]{article}
\def\epsfsize#1#2{1.0#1}
\epsfverbosetrue
\input amssym.def       
\input amssym.tex       

\def\goth{\frak}          
\def\double{\Bbb}

\def\rr{{\double R}}

\def\aa{{\cal A}}

\def\gg{{\goth g}}        
\def\hh{{\cal H}}

\def\aa{{\cal A}}
 
\def\hh{{\cal H}}

\def\t{{\rm tr}\,}

\def\bb{\begin{eqnarray}}
\def\ee{\end{eqnarray}}
\def\eee{\nonumber\end{eqnarray}}
\def\pp{\pmatrix}
\def\qq{\quad}
\def\del{\partial}

\begin{document}

\hsize 17truecm
\vsize 24truecm
\font\twelve=cmbx10 at 13pt
\font\eightrm=cmr8
\baselineskip 18pt

\begin{titlepage}

\centerline{\twelve CENTRE DE PHYSIQUE THEORIQUE}
\centerline{\twelve CNRS - Luminy, Case 907}
\centerline{\twelve 13288 Marseille Cedex 9}
\vskip 4truecm

\centerline{\twelve THE NONCOMMUTATIVE CONSTRAINTS }
\centerline{\twelve ON THE STANDARD MODEL \`a la CONNES}

\bigskip

\begin{center}
{\bf Lionel CARMINATI}
\footnote{ and Universit\'e de Provence\\
carminati@cpt.univ-mrs.fr \qq\qq 
\qq iochum@cpt.univ-mrs.fr \qq\qq
\qq schucker@cpt.univ-mrs.fr} \\
\bf Bruno IOCHUM$^{1}$\\
\bf Thomas SCH\"UCKER $^{1}$
\end{center}

\vskip 2truecm
\leftskip=1cm
\rightskip=1cm
\centerline{\bf Abstract} 

\medskip

Noncommutative geometry applied to the standard
model of electroweak and strong interactions was
shown to produce fuzzy relations among masses and
gauge couplings. We refine these relations and
show then that they are exhaustive.

\vskip 1truecm
PACS-92: 11.15 Gauge field theories\\ 
\indent
MSC-91: 81E13 Yang-Mills and other gauge theories 
 
\vskip 2truecm

\noindent April 1996
\vskip 1truecm
\noindent CPT-96/P.3307\\
\noindent hep-th/9604169
 
\vskip1truecm

 \end{titlepage}

\section{Introduction}
Connes' geometric version of the standard model
\cite{connes} needs no further introduction
\cite{intro}. For the physicist, its most interesting
feature is the explanation of spontaneous symmetry
breaking. Starting from the fermionic mass matrix and
 `noncommutative gauge couplings', this explanation
produces the bosonic mass matrices, spin 0 and 1, and
the ordinary gauge couplings.
Recall that the ordinary gauge couplings  $g_i, i {\in \{2,3 \}}$,
parameterize the most general invariant scalar
product on the Lie algebra $\gg$, e.g.,
\bb (X,X'):=\frac{2}{g_i^2}\,\t(X^*X'),\qq
X,X'\in {\rm su}(i). \eee
In noncommutative geometry, the Lie algebra $\gg$
is contained in the involution algebra $\aa$,
$\gg=\left\{X\in\aa,\ X^*=-X\right\}$ and the
invariant scalar product is constructed from the
fermion representation $\rho$, which now is a
representation of $\aa$ on a Hilbert space $\hh$,
\bb (a,a'):=\t(z\rho(a)^*\rho(a')),\qq a,a'\in\aa.\eee
The {\it noncommutative gauge coupling} $z$ is a positive
matrix on
$\hh$ that commutes with
$\rho(\aa)$ and with the fermionic mass matrix. $z$
unifies ordinary gauge couplings and boson masses.
 In the standard model, $z$ contains
 six positive numbers $x,\  y_1,\ y_2,\ y_3,\ \tilde x,\ 
\tilde y$ and the boson masses and gauge couplings as functions of these six numbers are \cite{ks}:
\bb 
m^2_W&=&\frac{xq+y_1m_e^2+y_2m_\mu^2+y_3m_\tau^2}
{3x+y_1+y_2+y_3},\label{W}\\\cr 
m^2_H&=&\frac{xr^2 +3(y_1m_e^4+y_2m_\mu^4+y_3m_\tau^4)}
{xq+y_1m_e^2+y_2m_\mu^2+y_3m_\tau^2}
 -({xq+y_1m_e^2+y_2m_\mu^2+y_3m_\tau^2}) \label{H} \\
&&\hspace{100pt}\left(\frac{1}{3x+y_1+y_2+y_3}+\frac{1}{3x+(y_1+y_2+y_3)/2}\right),\nonumber \\ 
g^{-2}_1&=&3x+\frac{2}{3}\,\tilde x+\frac{1}{2}\,
(y_1+y_2+y_3)+\frac{3}{2}\,\tilde y,\label{g1}\\
 g_2^{-2}&=&{3x+y_1+y_2+y_3},\label{g2}\\
 g^{-2}_3&=&4\tilde x.\label{g3}
\ee
Here, we have  denoted the mass of a particle $p$ by $m_p$ and put,
\bb
q&:=&m_t^2+m_b^2+m_c^2+m_s^2+m_u^2+m_d^2 \, \nonumber \\
r^2&:=&3(m_t^4+m_b^4+m_c^4+m_s^4+m_u^4+m_d^4) \nonumber \\
& &\qq+2\left[ (m_um_d|V_{ud}|)^2+(m_um_s|V_{us}|)^2+(m_um_b|V_{ub}|)^2  \right. \nonumber \\
& & \qq \qq \qq+\ (m_cm_d|V_{cd}|)^2+(m_cm_s|V_{cs}|)^2+(m_cm_b|V_{cb}|)^2 \nonumber \\
& &\qq \left. \qq \qq +\ (m_tm_d|V_{td}|)^2+(m_tm_s|V_{ts}|)^2+(m_tm_b|V_{tb}|)^2 \right],
\eee
and the $V_{..}$ are the Cabbibo-Kobayashi-Maskawa mixings.

In the ordinary formulation, the standard model has
18 positive input parameters: the three gauge
couplings, $g_1,\ g_2,\ g_3$, the $W$ and $H$ masses,
three lepton and six quark masses, and four angles contained in
the unitary Cabbibo-Kobayashi-Maskawa matrix $V$.

In its geometric formulation, there are 19 positive input
parameters, the 9 fermionic masses and 4 mixing angles
and 6 parameters from the noncommutative gauge
coupling. In order to derive the constraint equations
for $m_W,\ m_H,\ g_1,\ g_2,\ g_3$, one has to
distinguish several cases in terms of the 13 {\it
independent} parameters of the fermionic mass matrix.
The equations (\ref{W}-\ref{g3}) apply to the case
where the Cabbibo-Kobayashi-Maskawa matrix is
non-degenerate, i.e. not block diagonal up to
permutations of basis elements. In physical terms, this
means that there are no simultaneous mass and weak
interaction eigenstates. 

The following abbreviations will be useful:

\[\begin{array}{lll}
e:=m_{e}^2,& \mu:=m_{\mu}^2,& \tau:=m_{\tau}^2,\\ \\
t:=m_{t}^2,& b:=m_{b}^2,& c:=m_{c}^2,\qq..., \\ \\
W:=m_{W}^2,& H:=m_{H}^2.
\end{array}\]

Now, we are in a more symmetric situation of five equations for 
$m_W, m_H, g_1, g_2, g_3$ as functions of five unknowns 
{\it noncommutative gauge parameters}
$y_1,\ y_2,\ y_3,\ \tilde x,\ \tilde y$ 
and five {\it effective parameters}
$q,\ r,\ e,\ \mu,\ \tau$.

Our task is to describe the open subset of the five
dimensional space of $(m_W, m_H, g_1, g_2, g_3)$ that
is the image under equations (\ref{W}-\ref{g3}) 
of the five positive noncommutative gauge parameters. 
Of course this image varies with the effective parameters.
 Again, we have to distinguish cases in terms
of the five effective parameters  $q, r, e,\mu,
\tau$. 

Here, we treat only one simple case given
by the following hierarchies,
\begin{equation}
\left. \begin{array}{rcclcl}
e&<&\mu&<&\tau &< \qq\qq W, \\ \\
u+d&<&{\rm min}\{c,s\}&<&(1+\epsilon)^{-1}{\rm max}\{c,s\},\\ \\
 c+s+{\rm min}\{c,s\}&<&{\rm min}\{t,b\}&<&(1+\epsilon)^{-1}{\rm max}\{t,b\},
\end{array} \right\}
\label{Hierarchie} \end{equation}
where $\epsilon:=1-
\min\{|V_{tb}|^2,|V_{cs}|^2,|V_{ud}|^2\}$ measures the
deviation of the Cabbibo-Kobayashi-Maskawa matrix
from the identity.\\
\noindent These hierarchies are simply used for getting the positivity 
of the following constant \cite{iks}
\bb C&:=&\frac{r^2-q^2}{3W^2}\,>\,0 \label{C>0}.\ee
Actually, we write $C$ under the form
\bb \frac{3}{2}\,CW^2&=&
[t^2+b^2+c^2+s^2+u^2+d^2] \label{cterms} \\ 
&&\qq+\ [tb\,(|V_{tb}|^2-1)+cs\,(|V_{cs}|^2-1)+
ud\,(|V_{ud}|^2-1)]\cr 
&&\qq +\ [us\,|V_{us}|^2+ub\,|V_{ub}|^2+cd\,|V_{cd}|^2
+cb\,|V_{cb}|^2+td\,|V_{td}|^2+ts\,|V_{ts}|^2]\cr 
&&\qq -\ [(t+b)(c+s+u+d)+(c+s)(u+d)].\eee
In (\ref{cterms}), a lower bound of the second term of the right-hand side is 
$-\epsilon(tb+cs+ud)$, 0 for the third term and  $-[(t+b)\min\{t,b\}
+(c+s)\min\{c,s\}$ for the last one. According to the
definition of $\epsilon$, 
$3\,CW^2/2>u^2+d^2-\epsilon \,ud>0$. 

\section{Fuzzy relations for the masses and coupling constants}
Since the previous hierarchies (\ref{Hierarchie}) are experimentally 
true (cf the appendix), this hypothesis is not restrictive 
and we have the following
\begin{description}
\item{\bf Theorem.}
{\it 
\qq Assume (\ref{Hierarchie}): the heaviest lepton $\tau$ is lighter than the W 
and there is a hierarchy between quarks and mixings. Then, the image, in the five
dimensional space $(m_W, m_H, g_1, g_2, g_3)$, of the six strictly positive 
noncommutative gauge parameters 
$x,$ $y_1,$ $y_2,$ $y_3,$ $\tilde x,$ $\tilde y$, is characterized 
by the following five inequalities,
\bb 
\tau &< &\qq W \qq < \ q/3\,, \label{WI} \\
H_{min}(W) &<& H(W) \ < \ H_{max}(W)\,, \label{HI} \\
{\rm sin}^2\theta_w &<& \frac{2}{3}\,\left(1+\frac{W-\tau}
{q-3\tau}+(\frac{g_2}{3g_3})^2\,\right)^{-1}\,. \label{SIN}
\ee
The saturated bounds are given by
\bb H_{max}(W) := \frac{r^2-9e^2}{q-3e}\ -\ \frac{(r^2-3qe)e}{q-3e}\,\frac{1}{W}
  \ -\ \frac{3q+3W-12e}{q+3W-6e} W\, ,\label{Hmax}\ee
\bb H_{min}(W) := \frac{r^2-9\tau^2}{q-3\tau}\ -\ \frac{(r^2-3q\tau)\tau}{q-3\tau}\,\frac{1}{W}
 \ -\ \frac{3q+3W-12\tau}{q+3W-6\tau} W \,. \ee
In particular, 
\bb
H_{max}(W)-H_{min}(W)=(\tau-e)\,(q-3W)\, 
\left[\frac{r^2-3q(e+\tau)+9e\tau}{(q-3e)(q-3\tau)} \,\frac{1}{W}\right. \nonumber \\
 \left. +\  \frac{6W}{(q+3W-6e)(q+3W-6\tau)} \right] \,. \label{deltaH}
\ee
Note that the $3$ in $(\ref{WI})$ is the number of generations and that the 
intermediate lepton $\mu$ does not appear in these formulae.
}
\end{description}

This is the first time that we see a mass relation
affected by a small conceptual uncertainty. We call
it a {\it fuzzy mass relation} \cite{iks}. \\
{\it Proof:} Inequalities (\ref{WI}) follow immediately  from equation (\ref{W}).

\noindent The proof of (\ref{HI}) is more involved. Since the equations (\ref{W}-\ref{H}) 
are homogeneous in the $x, y_1,y_2,y_3$ variables, we will assume temporarily
\bb 3x = 1.\eee  As in \cite{iks}, we introduce two variables:
\bb 
X&:=&1+\sum_{j=1}^3y_j,\nonumber \\
Y&:=&\alpha_0^2+\sum_{j=1}^3\alpha_j^2y_j.
\eee with the following abbreviations
\bb \alpha_0&:=&q/3W,\qq\qq 
\alpha_1\ :=\ e/W,\qq\qq
\alpha_2\ :=\ \mu/W,\ \qq\qq
\alpha_3\ :=\ \tau/W.\eee
The hierarchy (\ref{Hierarchie}) and (\ref{WI}) imply
\bb \alpha_1 < \alpha_2<\alpha_3<1<\alpha_0.\eee
In terms of X and Y, the mass relations (\ref{W}-\ref{H}) read:
\bb
\frac{H}{W}\,+1&=&\frac{C}{X}\,+3\,\frac{Y}{X}\,
-2\,\frac{X}{1+X},\label{xhw}\\
X&=&\alpha_0+\sum_{j=1}^3\alpha_jy_j \label{xx}.\ee
It is convenient to define 
\bb
X_j&:=& \frac{\alpha_0-\alpha_j}{1-\alpha_j} \,, \nonumber \\
Y_j&:=& \beta_jX_j \,, \nonumber  \\
\beta_j&:=&\alpha_0+\alpha_j-\alpha_0\alpha_j \,, \qq j{\in \{1,2,3\}}.
\eee
Recall the following result of \cite{iks} and its proof for completeness:

\begin{description}
\item{\bf Lemma 1.} 
{\it
$D:=\left\{y=(y_1,...,y_3)/\ y_j>0,\ 
\sum_{j=1}^3(1-\alpha_j)y_j=\alpha_0-1\right\}$
is a convex open set in $\rr^3$. Moreover, on D, the variables
 X and Y are independent and satisfy 
$X{\in ]X_1,X_3[}, \, Y{\in]Y_1,Y_3[}$.
}
\end{description}
{\it Proof.}
\noindent  $D$ is convex and bounded: Indeed, for $j{\in \{1,2,3\}}$, 
we have
\bb (1-\alpha_j)y_j<\,\sum_{j=1}^3(1-\alpha_j)y_j=
\alpha_{0}-1,\eee
 and
$0<y_j<(\alpha_{0}-1)(1-\alpha_j)^{-1}.$
Let 
\bb A_1:=(X_1-1,0,0),\qq A_2:=(0,X_2-1,0) {\rm  \qq and\qq } A_3:=(0,0,X_3-1).\eee
Clearly, the $A_j$ are in the
closure of $D$ and $D$ is the interior of the convex
envelope of the vectors $A_j$: every $y=(y_1,y_2,y_3)\in D$
can be written as
\bb y=\sum_{j=1}^3\lambda_j A_j \qq{\rm with}\qq
\lambda_j:=\,\frac{1-\alpha_j}{\alpha_0-1}\ 
y_j>0\qq {\rm
and}\qq\sum_{j=1}^3\lambda_j=1\eee 
because of the constraint (\ref{xx}). Therefore
\bb X=1+\sum_{j=1}^3y_j=\sum_{j=1}^3\lambda_j
\left(1+\frac{\alpha_0-1}{1-\alpha_j}\right)
=\sum_{j=1}^3\lambda_j\,\frac{\alpha_0-\alpha_j}
{1-\alpha_j},\eee
and as $(\alpha_0-\alpha)/(1-\alpha)$ is an
increasing function of $\alpha$,
\bb \frac{\alpha_0-\alpha_1}{1-\alpha_1}\,<X<
\frac{\alpha_0-\alpha_3}{1-\alpha_3}.\eee
Similarly, we obtain the bounds on $Y$,
\bb
Y=\alpha_0^2+\sum_{j=1}^3\alpha_j^2y_j=\sum_{n=1}^3\lambda_n
\left(\alpha_0^2+
(\alpha_0-1)\frac{\alpha_n^2}{1-\alpha_n}\right)
\eee 
by noting that $\alpha^2/(1-\alpha)$ is increasing
in $\alpha$:
\bb \alpha_0^2+(\alpha_0-1)\,\frac{\alpha_1^2}
{1-\alpha_1}\,<Y<
\alpha_0^2+(\alpha_0-1)\,\frac{\alpha_N^2}
{1-\alpha_3}.\eee
In particular, $X_2 {\in ]X_1,X_3[} \ {\rm  and }\  Y_2 { \in]Y_1,Y_3[}$.\\
The independence of $X$ and $Y$ follows from a
non-vanishing functional determinant.
Solving the
constraint,
\bb y_3=-\,\frac{1-\alpha_0}{1-\alpha_3}\,
-\,\frac{1-\alpha_1}{1-\alpha_3}\,y_1
 -\,\frac{1-\alpha_2}{1-\alpha_3}\,y_2,\label{y_3}\ee
 we eliminate $y_3$:
\bb X
&=&\,\frac{\alpha_0-\alpha_3}{1-\alpha_3}\,+
\,\frac{\alpha_1-\alpha_3}{1-\alpha_3}\,y_1+
\,\frac{\alpha_2-\alpha_3}{1-\alpha_3}\,y_2,\cr \cr 
Y&=&\left(\alpha_0^2-\alpha_3^2\,
\frac{1-\alpha_0}{1-\alpha_3}\right)\,+
\left(\alpha_1^2-\alpha_3^2\,
\frac{1-\alpha_1}{1-\alpha_3}\right)\,y_1+
\left(\alpha_2^2-\alpha_3^2\,
\frac{1-\alpha_2}{1-\alpha_3}\right)\,y_2,
\eee
and compute the functional determinant
\bb \det\,\pp{
\del X/\del y_1&\del X/\del y_2\cr 
\del Y/\del y_1&\del Y/\del y_2}\,=\,
\frac{(\alpha_1-\alpha_2)(\alpha_2-\alpha_3)
(\alpha_3-\alpha_1)}{1-\alpha_3}\,{\neq 0}.\eee
ending the proof of the lemma.

The next lemma characterizes the domain $D$ as function of the
variables $X$ and $Y$.

\begin{description}
\item{\bf Lemma 2.}
{\it
  Let T be the map from $\rr^3$ to $\rr^2$ defined by 
$T(y_1,y_2,y_3):=(X,Y)$. Then, the image T(D) is the interior of 
the triangle delimited by the points $T(A_j)=(X_j,Y_j),  j{\in \{1,2,3\}}$. 
}
\end{description}

{\it Proof:} Since $y_3$ is positive, (\ref{y_3}) implies
\bb 0<y_2<-\frac{1-\alpha_1}{1-\alpha_2}\,
y_1 + \frac{\alpha_0-1}{1-\alpha_2}\,. \eee
This upper bound being a line in the ($y_1,y_2$) plane, 
the projection of $D$ on this plane is contained
 in the triangle defined by 
the points $(X_1-1,0)$, $(0,X_2-1)$ and (0,0). These points are nothing
 but the projection of $A_1, A_2, A_3$ which are in the closure
 of $D$. The projection on the plane preserves 
convexity and the previous Lemma yields the result because 
\bb
X_{T(A_j)}&=&1+X_j-1=X_j \nonumber \\
Y_{T(A_j)}&=&\alpha_0^2 + \alpha_j^2(X_j-1) =(\alpha_0-\alpha_j)(\alpha_0+\alpha_j)+\alpha_j^2X_j \nonumber \\ 
&=&(1-\alpha_j)X_j(\alpha_0+\alpha_j)+\alpha_j^2X_j=(\alpha_0+\alpha_j
-\alpha_0\alpha_j)\,X_j=Y_j.
\eee

Thanks to (\ref {xhw}), we need to control the function
\bb f(X,Y):=\,\frac{C}{X}\,+3\,\frac{Y}{X}\,-2\,\frac{X}{1+X} \, .\eee
$C$ being positive by (\ref{C>0}), $f$ is decreasing in $X$ and increasing in $Y$. 
So the minimum and maximum of $f(X,Y)$ for $(X,Y)\in T(D)$ 
 lie on the three segments [$T(A_j$),$T(A_k$)], $j\neq k$, 
which are the boundaries of $T(D)$. The points of 
these segments have coordinates of the form $(X, a_{jk}X+b_{jk})$ where 
$a_{jk}, b_{jk}$ are real numbers. The derivative of 
$g_{jk}(X):= f(X,a_{jk}X+b_{jk})$ is 
$g_{jk}'(X)= -(C+3b{jk})(1+X^2)^{-1}-(1+X)^{-2}$
and the functions $g_{jk}$ will be decreasing if $b_{jk}$
 is positive which is the case as proved in the next lemma,
 because $b_{jk}=(\alpha_0-\alpha_j) (\alpha_0-\alpha_k)$.
 This shows that
\bb
{\rm max} \{ f(X,Y)\ |\  (X,Y) \in D\}&=& {\rm max}\{f(X,Y)\ |\  (X,Y) \in [T(A_1),T(A_3]\} \nonumber \\
 &=& {\rm max} \{g_{13}(X)\ | \ X\in [X_1,X_3]\} = g_{13}(X_1) \nonumber \\
&=& f(X_1,Y_1),\nonumber \\
{\rm min} \{f(X,Y)\ |\ (X,Y)\in D\}&= &g_{13}(X_3) = g_{23}(X_3) = f(X_3,Y_3).
\eee
Now, by (\ref {xhw}), 
\bb 
H_{max} &=&W(3\beta_1 +\frac{C}{X_1}\, -\frac{2X_1}{1+X_1}\, -1) \\ \nonumber
&=& q+3e-\frac{qe}{W}+3\,C\, \frac{W-e}{q-3e} W \ -\ \frac{3q+3W-12e}{q+3W-6e} W
\eee
 yielding (\ref {Hmax}). This proves (\ref{Hmax}-\ref {deltaH}).

\noindent Note that 
\bb H_{min} = W(3\beta_3 +\frac{C}{X_3}\, -\frac{2X_3}{1+X_3}\, -1)\eee
is positive because $-2X(1+X)^{-1}-1 >-3$ for any positive $X$ and 
$\beta_3 = \alpha_0-\alpha_3(\alpha_0-1)>\alpha_0-(\alpha_0-1)=1$.

\begin{description}
\item{\bf Lemma 3.}
{\it
  The equation of the line passing through the points $T(A_i)$ and $T(A_j)$
in the (X,Y) plane is $\, Y=(\alpha_i+\alpha_j-\alpha_i\alpha_j)\,X
 + (\alpha_0-\alpha_i)(\alpha_0-\alpha_j)$.
}
\end{description}

{\it Proof:} This line is $Y=(Y_j-Y_i)(X_j-X_i)^{-1}\,X
+(Y_iX_j-Y_jX_i)(X_j-X_i)^{-1}$ and
\bb
\frac{Y_iX_j-Y_jX_i}{X_j-X_i}&=&(\beta_i-\beta_j) \frac{X_iX_j}{X_j-X_i}\nonumber \\
&=&(\alpha_j-\alpha_i)(\alpha_0-1)\,\frac{(\alpha_0-\alpha_j)\,(\alpha_0-\alpha_i)}
{(\alpha_0-\alpha_j)(1-\alpha_j)-(\alpha_0-\alpha_i)(1-\alpha_i)} 
\nonumber \\
&=&(\alpha_0-\alpha_i)\,(\alpha_0-\alpha_j)\,.
\eee
Moreover, the slope is
$(Y_j-Y_i)(X_j-X_i)^{-1}= (\alpha_i+\alpha_j-\alpha_i\alpha_j)\,$.
\bigskip

 To include the coupling constants, equations (\ref{g1}-\ref{g3}),
we remark that the $W$ and Higgs masses are
homogeneous in $x,\ y_1,\ y_2,\ y_3$ and independent
of $\tilde x, \tilde y$. Consequently, the image under
equations (\ref{W},\ref{H},\ref{g2},\ref{g3}) is a cylinder with  $g_2>0$, $g_3>0$ 
with basis  given by the 
inequalities (\ref{WI},\ref{HI}) and shown in Figure \ref{fig:Hmax and Hmin}.
 At this point, $x$ is arbitrary positive as  $\tilde x$ and so are $g_2$ and $g_3$.
To solve the last constraint (\ref{g1}), we write
\bb (y_1+y_2+y_3)\,e\,<\,y_1e+y_3\tau \,<\,(y_1+y_2+y_3)\,\tau,\eee
and from (\ref{W},\ref{g2})
\bb (\frac{q}{3}-W)g_{2}^{-2}\,=\,(y_1+y_2+y_3)\,\frac{q}{3} - (y_1e+y_3\tau),\ee
we obtain two optimal inequalities
\bb
\frac{q/3-W}{q/3-e}\,g_{2}^{-2}\,<\,y_1+y_2+y_3\,<\,\frac{q/3-W}{q/3-\tau}\,g_{2}^{-2}.
\eee
This solves the constraint (\ref{g1}) on $g_1$:
\bb
\frac{1}{2}\,g_2^{-2}\left(1+\frac{W-\tau}{q/3-\tau}
\right)+\frac{1}{6}\,g_3^{-2}+\frac{3}{2}\,\tilde y
\,<\,g_1^{-2}\,<\,
\frac{1}{2}\,g_2^{-2}\left(1+\frac{W-e}{q/3-e}
\right)+\frac{1}{6}\,g_3^{-2}+\frac{3}{2}\,\tilde
y.\eee
Since $\tilde y$ is an arbitrary positive number, we finally get
\bb \frac{1}{2}\,g_2^{-2}\left(1+\frac{W-\tau}
{q/3-\tau}\right)+\frac{1}{6}\,g_3^{-2}\,<\,g_1^{-2}
\eee
which is nothing else but (\ref{SIN}) with 
sin$^2\theta_w=g_2^{-2}(g_1^{-2}+g_2^{-2})^{-1}$ and the theorem is proved.

\bigskip

\noindent {\bf Problem}: It would be interesting to get the Theorem without the hierarchy (\ref{Hierarchie}).

\section{Physical consequences}

\begin{figure}[hbt]
\hspace{2cm}
\def\epsfsize#1#2{0.85#1}
\epsfbox{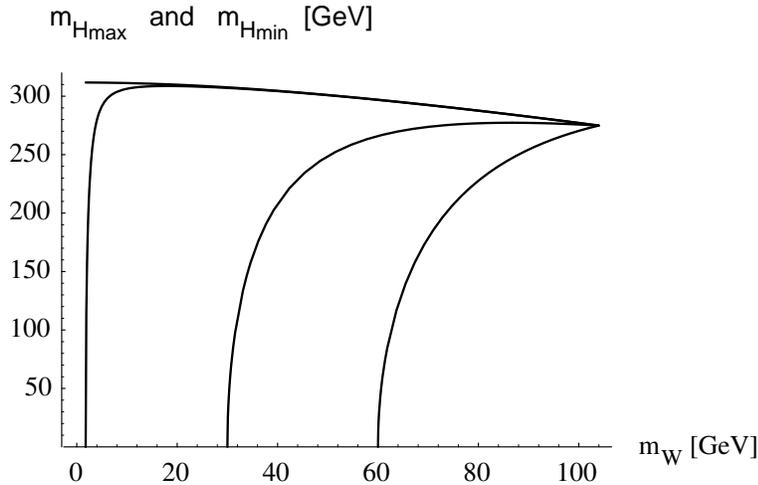}
\caption{$m_{H_{max}}$ and $m_{H_{min}}$ as function of $m_W$ for $m_{\tau} = 1.8, 30 {\rm \  and}\  60$ GeV}
\label{fig:Hmax and Hmin}
\end{figure}

The inequality (\ref{WI}) is
\bb m_W < \sqrt{\frac{q}{3}} = 104 \, {\rm \, GeV}. \eee
The three inequalities (\ref{HI}-\ref{SIN}) deserve a few 
graphic representations. Figure \ref{fig:Hmax and Hmin} shows the allowed domain 
for the Higgs mass as a function of $m_W$ with  $m_{\tau}$ as a parameter. The upper curve is  $m_{H_{max}}$
which is independent of $m_{\tau}$. 
All parameters not explicitly mentioned in a figure or its caption are set to their experimental
central values e.g. in Figure \ref{fig:Hmax and Hmin}, $m_t = 180$ GeV. For the experimental values 
$m_W = 80$ GeV and $m_{\tau} = 1.8$ GeV, the allowed interval for the Higgs mass collapses
in Figure \ref{fig:Hmax and Hmin}. Indeed, this {\it conceptual} uncertainty, 'fuzziness', is 
\bb
m_{H_{max}}-m_{H_{min}}\, = 34\, {\rm \, MeV}.
\eee
The fuzziness is controlled by the $\tau$ mass :
\bb
\frac{m_{{H}_{max}}-m_{{H}_{min}}}{m_{{H}_{max}}+m_{{H}_{min}}}\,\sim\,
\frac{m_\tau^2-m_e^2}{m_t^2}\, \sim \, 10^{-4} \qq {\rm \ at} \  m_W = 80 \ {\rm GeV} \label{dif}
\eee
 and disappears at the upper bound $m_W = 104$ GeV since
\bb H_{min}(\frac{q}{3}) = H_{max}(\frac{q}{3}) = 
\frac{r^2}{q}-\frac{2q}{3} = (275 \,{\rm GeV})^2.\eee
Note that this value of $m_H$ is independent of the lepton masses. 

\begin{figure}[hbt]
\hspace{2cm}
\epsfbox{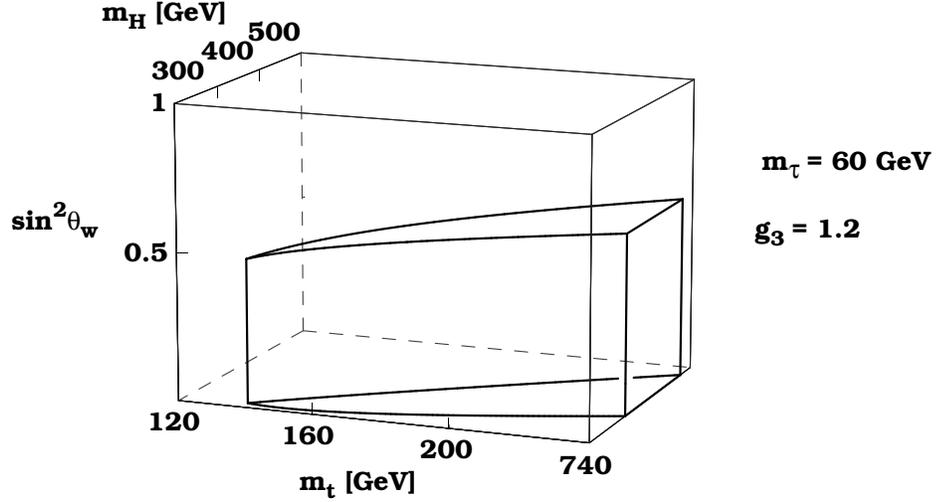}
\caption{Allowed parameter domain, unrealistic }
\label{fig:Allowed parameter domain}\end{figure}
\begin{figure}[hbt]
\hspace{2cm}
\epsfbox{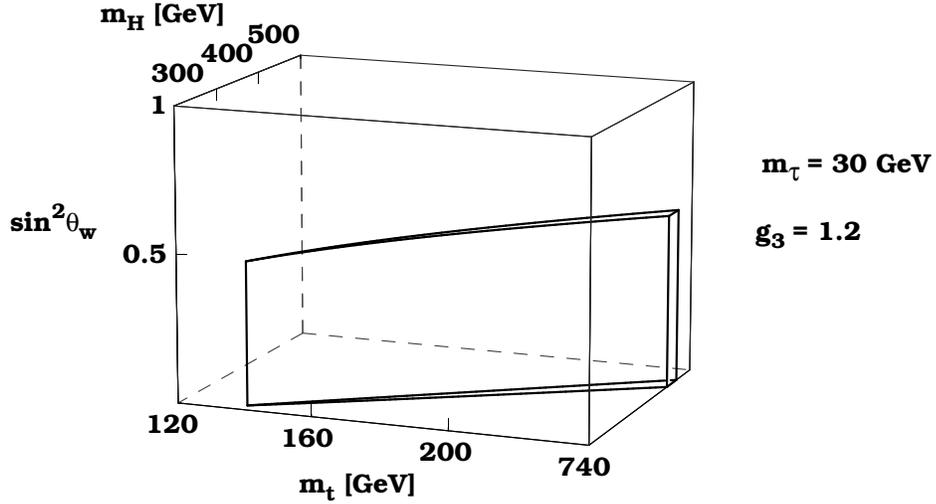}
\caption{The real collapse}
\label{fig:The real collapse}
\end{figure}
\begin{figure}[hbt]
\hspace{2cm}
\epsfbox{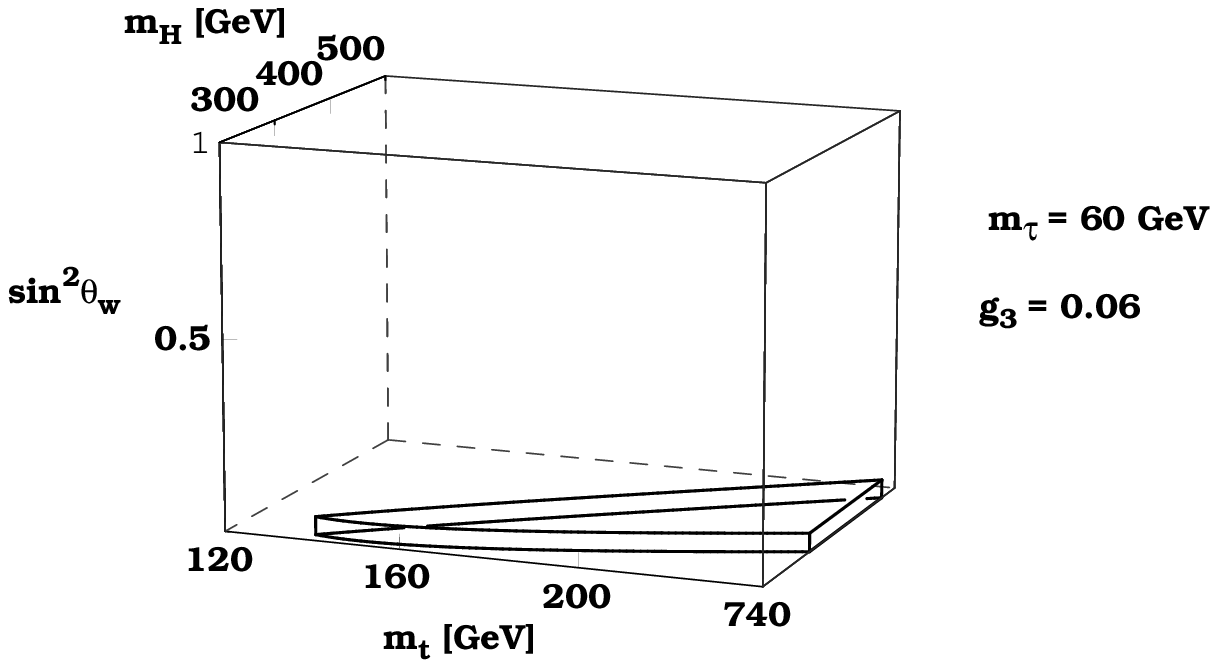}
\caption{An academic collapse}
\label{fig:An academic collapse}
\end{figure}
\vskip 0.7truecm
 
In any case, the {\it experimental} uncertainties on the masses, completely drown the fuzziness. 
Since, today, the major experimental uncertainty is on the top mass, $\pm 12$ Gev, 
it is worth to represent the fuzziness as function of $m_t$ with $m_{\tau}$ as parameter. 
Figures \ref{fig:Allowed parameter domain} and \ref{fig:The real collapse} illustrate again the mentioned mass collapse.
To incorporate inequality (\ref{SIN}), we include in these figures $\sin^2 \theta_w$ and $g_3$. 
A second collapse in $\sin^2 \theta_w$  is ploted in Figure \ref{fig:An academic collapse}.

Neglecting the fuzziness with respect to experimental accuracy, the inequalities (\ref{WI}-\ref{SIN}) 
reduce to \cite{ks}:
\bb
m_e&<&m_W\, \lesssim \,m_t/\sqrt 3, \nonumber \\ \cr 
 m_H^2&\approx&3\,\frac{(m_t/m_W)^4+2(m_t/m_W)^2-1}
{(m_t/m_W)^2+3}\,\ m_W^2 \label{HH}, \\\cr 
\sin^2\theta_w& \lesssim &\frac{2}{3}\,\frac{1}
           {1+m_W^2/m_t^2+(g_2/3g_3)^2}\label{SS}.
\ee
Note that $m_W > m_e$ does not use the hierarchy (\ref{Hierarchie}).

The last inequality has two physical consequences: if
we know the $W$ and fermion masses, then, the weak
angle is constrained by $\sin^2\theta_w<0.54$ (Figure \ref{fig:An academic collapse}).
Recall the experimental values, $\sin^2\theta_w=0.23$, $g_3=1.2$.
If we know the $W$ and fermion masses and the
electroweak couplings, then, the strong coupling
cannot be too weak: $g_3>0.17$ at the $Z$ mass.
At this point, the following fact \cite{RA} is intriguing: 
If we know the fermion representation under electroweak interactions, 
then, the strong interactions must be vectorlike. If not, the noncommutative 
generalization of Poincar\'e duality breaks down.

One of the attractive features of noncommutative
geometry in Yang-Mills theories is that gauge
couplings and boson masses are correlated.
Numerically, this unification is visible in the
inequality (\ref{SS}). Adding right-handed neutrinos
to the standard model \cite{gb} improves this bound
\bb\sin^2\theta_w<\frac{1}{2}\,
\frac{1}{1+g_2^2/(12g_3^2)}\,.\eee
However in this case, gauge couplings and masses decouple and 
also Poincar\'e duality breaks down.

\section{Conclusion}

The fuzzy mass relation for the Higgs raises the
question of stability under renormalization. We feel
that this question can only be answered by taking
seriously the revolution that noncommutative
geometry operates on spacetime. Spacetime becomes
fuzzy \cite{john}, just as phase space becomes fuzzy in
quantum mechanics. Let us try to explain this feeling
by an analogy with electrodynamics. Unifying
electricity and magnetism, Maxwell obtained an
expression for the speed of light in terms of the two
static coupling constants $\epsilon_0$ and $\mu_0$. His relation was confirmed
by already existing data and no-one really dared to
ask, what could be the meaning of an equation between
a quantity depending on the reference system and a
constant. Later, Einstein answered the question with
the help of Minkowskian geometry. This geometry was
already inherent in Maxwell's equations, but not
accepted by the community of physicists. Noncommutative
geometry tells us that the Higgs field with its
spontaneous symmetry breaking is only a magnetic
field and therefore the Higgs mass is fixed, fuzzily. 
This seems in contradiction with large renormalization
flow. However the origin of this flow is a small distance
divergence that ignores the new spacetime
uncertainty. In this context, quantum field theory has
to be redone \cite{dfr}. 

Meanwhile, we are looking
forward to the LHC verdict concerning equation
(\ref{HH}),
\bb m_H = 288 \pm22 {\rm \, GeV} \qq{\rm if} \qq m_t  = 180 \pm12 {\rm \, GeV}. \eee

\section{Appendix}

The present experimental
constraints \cite{data} on the 18 parameters of the
standard model are listed below. The gauge couplings are given at the
$Z$ mass and all masses are pole masses.
\[\left\{\begin{array}{rcl}
g_1&=&0.3575\pm 0.0001,\\
g_2&=&0.6507\pm 0.0007,\\
g_3&=&1.207\pm 0.026,
\end{array}\right.\]
\[\left\{\begin{array}{rcl}
m_e&=&0.51099906\pm 0.00000015\ {\rm MeV},\\
m_\mu&=&0.105658389\pm 0.000000034\ {\rm GeV},\\
m_\tau&=&1.7771 \pm 0.0005\ {\rm GeV},
\end{array}\right.\]
\[\left\{\begin{array}{rclrcl}
m_u&=&5\pm 3\ {\rm MeV},&m_d&=&10\pm 5\ {\rm MeV},\\
m_c&=&1.3\pm 0.3\ {\rm GeV},&m_s&=&0.2\pm 0.1\ {\rm GeV},\\
m_t&=&180\pm 12\ {\rm GeV},&m_b&=&4.3\pm 0.2\ {\rm GeV},
\end{array}\right.\]
\[\begin{array}{rcl}
m_W&=&80.22\pm 0.26\ {\rm GeV},\\
m_H&>&58.4\ {\rm GeV}.
\end{array}\]
The Cabbibo-Kobayashi-Maskawa matrix is a
unitary matrix
\bb V:=\pp{V_{ud}&V_{us}&V_{ub}\cr 
V_{cd}&V_{cs}&V_{cb}\cr 
V_{td}&V_{ts}&V_{tb}}\eee
 and the absolute values of its matrix elements are:
\bb \pp{
0.9753\pm 0.0006&0.221\pm 0.003&0.004\pm
0.002\cr 
0.221\pm 0.003&0.9745\pm 0.0007&0.040\pm 0.008\cr 
0.010\pm 0.006&0.039\pm 0.009&0.9991\pm 0.0004}.
\eee
For physical purposes, the 
Cabbibo-Kobayashi-Maskawa matrix can be
parameterized by three angles, $\theta_{12}$,  
$\theta_{23}$, $\theta_{13}$ and
one $CP$ violating phase $\delta$:
\bb V=\pp{
c_{12}c_{13}&s_{12}c_{13}&s_{13}e^{-i\delta}\cr 
-s_{12}c_{23}-c_{12}s_{23}s_{13}e^{i\delta}&
c_{12}c_{23}-s_{12}s_{23}s_{13}e^{i\delta}&
s_{23}c_{13}\cr 
s_{12}s_{23}-c_{12}c_{23}s_{13}e^{i\delta}&
-c_{12}s_{23}-s_{12}c_{23}s_{13}e^{i\delta}&
c_{23}c_{13}},\eee
with $c_{kl}:=\cos \theta_{kl}$, 
$s_{kl}:=\sin \theta_{kl}$.

 \end{document}